\documentclass[sigconf]{acmart}
\AtBeginDocument{%
  \providecommand\BibTeX{{%
    \normalfont B\kern-0.5em{\scshape i\kern-0.25em b}\kern-0.8em\TeX}}}

\setcopyright{acmcopyright}
\copyrightyear{2021}
\acmYear{2021}
\acmConference[TRAITS 2021]{The Road to a successful HRI: AI, Trust and ethicS (TRAITS) Workshop }{March 12, 2021}{Virtual}
\acmBooktitle{HRI 2021,
  March 09--11, 2021, Virtual}

\begin{document}

%\title{Sensing Trust through Speech}
\title{Speaking of Trust - Speech as a Measure of Trust}
%In Speech We Trust
%Speech: A Real-Time and Objective Trust Measure
%Speaking of Trust

%\author{ANONYMOUS}
%\email{anon@ymous.com}
%\orcid{XXXXXXX}
%\affiliation{%
%  \institution{University of Anon}
%  \city{XXX}
%  \country{XXX}
%}

\author{Ella Velner}
\email{p.c.velner@utwente.nl}
\affiliation{%
  \institution{University of Twente}
  \city{Enschede}
  \country{The Netherlands}
}

\author{Khiet P. Truong}
\email{k.p.truong@utwente.nl}
\affiliation{%
  \institution{University of Twente}
  \city{Enschede}
  \country{The Netherlands}
}

\author{Vanessa Evers}
\email{vanessa.evers@ntu.edu.sg}
\affiliation{
 \institution{Nanyang
Technological University}
 \city{Singapore}
 \country{Singapore}
}


\begin{abstract}
Since trust measures in human-robot interaction are often subjective or not possible to implement real-time, we propose to use speech cues (on what, when and how the user talks) as an objective real-time measure of trust. This could be implemented in the robot to calibrate towards appropriate trust. However, we would like to open the discussion on how to deal with the ethical implications surrounding this trust measure.
\end{abstract}



\maketitle

\section{Introduction}
Researchers in Human-Robot Interaction (HRI) are often interested in the trust people have in the robot. Understandable, since trust can be a prerequisite for relationship formation and self-disclosure \cite{Song2018HelpingTrust, Wheeless1977TheSelf-Disclosure}. Another reason for measuring trust, is to assess whether there is appropriate trust, meaning that the trust matches the capabilities of the robot \cite{Lee2004TrustReliance}. Overtrust can lead to complacency and dangerous situations, while undertrust can lead to underuse and suboptimal use of the robot's capabilities~\cite{Freedy2007MeasurementCollaboration,Lee2004TrustReliance}.
At the moment trust is mostly measured by either (self-reported) subjective measures (e.g., questionnaires) that might be influenced by people-pleasing behavior of the subjects, or behavioral measures that constrain the experiment design (e.g., trust games). Moreover, these measures are most often not measured real-time, so it is not possible to achieve optimal trust balance \emph{while} interacting with the robot. Hence, we propose to explore how objective and real-time sensing of trust can be achieved. 

Building on the knowledge that speech carries signals of emotions and attitudes \cite{Fernandez2005ClassicalSpeech, Kim2017TowardsLearning, Scissors2008LinguisticCMC, TerMaat2010HowAgent}, we propose to use the user's speech to measure the trust they have in the robot, by looking at \emph{what} they say, \emph{when} they say it and \emph{how} they say it. We are especially interested in how this could help the field of \emph{child}-robot interaction, since subjective measures are even more problematic for this user group \cite{Belpaeme2013Child-RobotChallenges}. In this short paper we will discuss how we envision a real-time measure of trust in human-robot interaction, but also address the ethical implications surrounding real-time trust calibration. 

\section{Trust Measures}
To measure trust, we can distinguish between three types of measures: subjective, active objective and passive objective measures. We refer to subjective measures as self-reports from the user. While these are widely used by researchers (\cite{Jian2000FoundationsSystems, Schaefer2016MeasuringScale-HRI, Yagoda2012YouScale}), they are administered after the actual interaction, hence they cannot be used as a real-time measure for the robot to sense the trust level during the interaction. Furthermore, they rely on the honesty and understanding of the user. While this is already somewhat problematic for adults, when children are involved, questionnaires could be quite unreliable. Children are known people-pleasers and therefore it can be hard to rely on their answers \cite{Belpaeme2013Child-RobotChallenges}.

When interaction designers provoke the user to make a choice during the interaction that would reflect their trust in the robot, this is considered an active objective measure. A classic example of such a measure is the investment game \cite{Berg1995TrustHistory}. In this game a truster invests money (or tokens) in a trustee (this investment is the measure of trust), and they later lose, earn or keep the money they invested. While these measures are not dependent on the self-report of users, and thus more objective \cite{Lewis2018TheInteraction}, they constrain the design of the interaction, since they must include these behavior-provoking scenarios. 

With a passive objective measure, we can monitor certain behaviors that are correlated with trust, without necessarily being constrained by certain behavior-provoking scenarios. Examples of such passive objective measures are the user's heart rate \cite{Gupta2019InVR}, the distance between the robot and the participant \cite{Babel2021SmallProximity}, and the words shared by the robot and the user \cite{Scissors2008LinguisticCMC}. Some of these could be measured by video observation, which, if automated, would still be an objective measure. However, if humans are involved in the annotation, it could become a less objective measure. We are especially interested in the automated types of measures, because these are not constraining the interaction design, and can be used real-time by the robot in an autonomous way during the interaction.

\section{Speech as a measure of trust}
Since many human-robot interactions use speech to communicate, it would be useful if the robot could use this to assess the trust level. We propose to look at dialog cues (what people say), interaction cues (when people say it) and vocal cues (how they say it). 

First, we can use what people say to measure trust. Scissors et al.~\cite{Scissors2008LinguisticCMC} already discovered that parties who trust each other often use the same \textbf{words}. It could also be interesting to look at \textbf{speech acts} that people use, since these have been previously associated with attitudes towards the conversation partner \cite{Barriere2017HybridInteractions}. Speech act classification can also be automated \cite{Liu2017UsingFramework}.
%what people say

Second, it would be interesting to look at when the user speaks. Elkins et al.~\cite{Elkins2012PredictingDynamics} describe a model where the agent takes the user's demographics and voice, and the \textbf{duration of the response} into account as a perception of trust. Furthermore, interpersonal attitudes, such as trust, can be reflected in the \textbf{turn-taking system} of the conversation \cite{Ravenet2015ConversationalInteractions}.
%when people say things

Third, we know that how people say things can show how they feel \cite{Kim2017TowardsLearning}. Waber et al.~\cite{Waber2015AResearch} found that trust is reflected in the amount of emphasis (a combination of \textbf{pitch}, \textbf{intensity} and \textbf{speech rate}) on words the truster uses, although it could be this is only reflected during first encounters or after trust violations. 
%how people say things

We will look at these cues simultaneously as we expect that they complement and relate to each other, which will give us a more complete and better understanding of how trust is expressed in speech. 

\section{Sense-Think-Act}
%explain how this measure of trust would be appropriate in the sense-think-act cycle. 

\begin{figure}
    \centering
    \includegraphics[width=7cm]{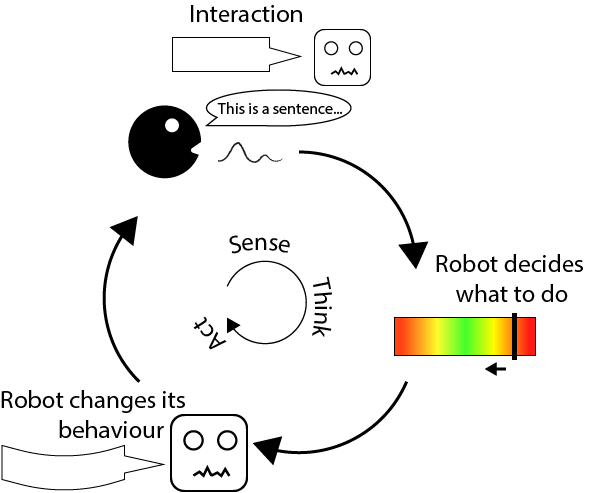}
    \caption{Using speech to integrate a sense-think-act cycle in the interaction}
    \label{fig:cycle}
\end{figure}

Measuring trust real-time and objectively would not only be beneficial to HRI researchers because of its objectiveness, but also to designers of HRI, that can incorporate this in a sense-think-act cycle, such as in Fig. \ref{fig:cycle}. An interaction should aim for appropriate trust, since overtrust can lead to complacency and dangerous situations, while undertrust can lead to underuse and suboptimal use of the robot's capabilities~\cite{Freedy2007MeasurementCollaboration,Lee2004TrustReliance}. This creates a spectrum of trust levels, where appropriate trust is found in the middle as the balance between overtrust and undertrust. When a user is trusting the robot too much, this could lead to dangerous situations, since the user would believe the robot to always act to the benefit of the user, while this could not be the case. Undertrust, on the other hand, could mean that the user would discard the robot, while it could have been trying to help the user. If the robot could \emph{sense} the trust a person has in it, it could \emph{think} about what the robot should do with this information (to try to lower or raise trust), and it could \emph{act} to reach appropriate trust balance \cite{Lee2004TrustReliance}.
If under- and overtrust could be avoided by calibrating towards appropriate trust, this could lead to more responsible interactions: the user needs to remain critical towards the robot, without discarding it entirely. Hence, a real-time measure of trust could play a part in creating more responsible human-robot interaction.

\section{Ethical implications}
Although using speech as a measure of trust has many advantages, as discussed above, its ethical implications should be addressed as well. In order to be able to extract trust from speech, the robot needs to record and analyze the voice of its users. Voice characteristics and speech can give insight in a user's personal information, such as their identity, personality and emotions \cite{Kroger2020PrivacyInference}, and hence, is a privacy-sensitive source. Consequently, before a user interacts with the robot, consent needs to be given, where the interaction designers can promise that the voice is only used for the benefit of this interaction. Moreover, the recorded data needs to be stored only when absolutely necessary (e.g., to improve the model) and with the utmost care by securing it with encryption and limited access \cite{Nautsch2019PreservingCharacterisation, Nautsch2019TheUnderstanding}. 

If our trust measure is applied within the sense-think-act cycle, more possible dangers come to mind. If trust can be calibrated towards appropriate trust, it could also be used to actually move away from appropriate trust. This way, robots could intentionally create overtrust, with all its consequences. For example, a company selling dubious products installs a robot to interact with potential buyers, and the robot increases their trust in it, so that when the robot says ``You'll definitely need this product'', the customer will buy it without thinking much about the possible risks \cite{vanderHeijden2003UnderstandingPerspectives}. Therefore, we should monitor that trust calibration is used towards responsible interaction. Furthermore, matching certain behavior of the robot to a trust level could induce stereotypes \cite{Hwang2019TowardsAgents}. For example, voice assistants use female voices because it is deemed more trustworthy, but this could be problematic due to a voice assistant always saying yes (giving the image of a submissive woman).

\section{Discussion}
In this short paper, we proposed to use speech as a real-time measure of trust. Trust is valuable to measure in an interaction, because it acts as a prerequisite for a fruitful interaction \cite{Ronfard2018PreschoolersAccuracy} and can be used to calibrate trust levels in real-time. Moreover, using speech cues as a measure would be an objective behavioral measure, without constraining the interaction. However, detecting trust from speech cues brings its ethical challenges. It would need privacy policies and, if implemented in a sense-think-act cycle, could have societal consequences.
We hope this paper opens a discussion on how we should balance these two sides of the same coin. Furthermore, the generalizability of speech as a measure of trust should be studied (e.g., does it also apply to human-human interaction?). We also want to acknowledge that trust might also be measurable by other modalities such as facial expressions, body posture, hand gestures, and eye gaze, however, speech is most often the main modality used and could potentially be of more wider use to different kinds of robots (also including voice assistants). Future research could investigate the role of other modalities and ways to measure trust objectively and real-time, possibly in combination with speech.

\bibliographystyle{ACM-Reference-Format}
\bibliography{main}
\end{document}